\documentclass{llncs}
\usepackage{times,epsfig,amssymb,latexsym,lscape}

\newcommand{\cpp}{\hbox{{\tt C++}}}
\newcommand{\para}{\vspace{0.05in}}
\newcommand{\pnt}{{\tt Point}}
\newcommand{\lin}{{\tt Line}}
\newcommand{\pln}{{\tt Plane}}
\newcommand{\cnt}{{\tt Constraint}}

\title{Computational Euclid}
\author{
 M.H. van Emden and B. Moa\\
    University of Victoria\\
    Victoria, B.C., Canada\\
}
\date{}
\institute{Research Report DCS-315-IR \\
           Department of Computer Science\\
           University of Victoria}

\begin{document}

\maketitle

\abstract{
We analyse the axioms of Euclidean geometry according to standard
object-oriented software development methodology. We find a perfect
match: the main undefined concepts of the axioms translate to object
classes.
The result is a suite of \cpp\ classes 
that efficiently supports the construction of complex geometric configurations.
Although all computations are performed in floating-point arithmetic,
they correctly implement as semi-decision algorithms
the tests for
equality of points, a point being on a line or in a plane,
a line being in a plane,
parallelness of lines, of a line and a plane, and of planes.
That is, in accordance to the fundamental
limitations to computability requiring
that only negative outcomes are given with certainty, while positive
outcomes only imply possibility of these conditions being true.
}

\section{Introduction}

We wrote a small class library to render
with computer graphics images of
the highly mathematical structures
created by the artist Elias Wakan \cite{wakan06}.
The \cpp\ classes we wrote will be extended
to output the type of description
required by a rendering package such as POV-Ray \cite{povRay06}.
We were surprised that this thoroughly practical enterprise
led us to two fascinating fundamental issues:
the limits of computability
and the abstract nature of axiomatic geometry.
Hence our title ``Computational Euclid''.

\para
To accommodate the limits of computability,
we use interval arithmetic.
We use it in such a way that when we pose the question
whether two straight lines intersect, 
the answer ``no'' has the force of a mathematical proof they do not.
The other possible answer is a box in 3-space, typically very small.
This answer means that the intersection is in this box,
\emph{if there is an intersection}.
This proviso is probably essential
because we sense that a decision procedure
for the intersection of two lines can be used
to implement a decision procedure
for the equality of any two real numbers,
which has been shown to be impossible by Turing \cite{trng37,aberth98}.
However, we have not pursued the details of such a problem reduction.

The other fundamental issue
is the abstract nature of an axiomatic approach to geometry.

Euclid is widely credited with inaugurating the axiomatic method in which
axioms contain references to undefined concepts of which the meaning is
only constrained by the axioms. However, we should not look to the Elements
for a literal embodiment of the axiomatic method.
It fell to Hilbert in 1902 \cite{hlbrt02} to cast these in a form that
is recognized today as an axiomatic treatment of the subject.

Lack of space prevents us here to go into a detailed analysis of the concepts
of Euclid's geometry.
Suffice it to say that Hilbert's formulation contains as undefined concepts,
among others, the following that we found useful in our work:
\emph{point},
\emph{line},
\emph{segment},
\emph{plane},
\emph{angle}.

In the modern conception of the axiomatic method these are undefined.
Their meaning is only constrained by the relations between them as
asserted by the axioms. In logic this is formalized by the axioms being
a theory, which, if consistent, can have a variety of models.
It is only the model that says what a point \emph{is}.
For example, in one type of model,
points, lines, and planes are solution spaces
of sets of linear equations in three variables.

What we find surprising is how well the way in which the axiomatic method,
as realized in formal logic, combines with the most widely accepted principles
of object-oriented software design \cite{wbww90}.
According to it, one looks for the \emph{nouns} in an informal specification
of the software to be written.
These are candidates for the \emph{classes} of an object-oriented program.

We used Hilbert's axioms as specification.
The recipe of \cite{wbww90} has, of course, to be taken with a grain of salt:
only the \emph{important} nouns are candidates for classes.
Usually, informal specifications contain a majority of not-so-important nouns.
To our delight,
we found that Hilbert's axioms
contain an unusually small number
of not-so-important nouns.

\section{The structure of our class library}

Of the nouns occurring in Hilbert's axioms,
\pnt, \lin, and \pln\ are a special subset.
They are special in the sense
that any unordered pair of these determine
an object in this trinity,
unless a specific condition prevails.
In the case of a point and a line determining a  plane,
the condition is ``unless the point is on the line''.
Note that these conditions are called \emph{predicates}
by some authors~\cite{effpred02,itar98}.
The table in Figure~\ref{pointsLinesPlanes}
summarizes the operations for all unordered pairs,
each with the attendant disabling condition.
\begin{figure*}
\begin{center}
\begin{tabular}{l|l}
\hline
 Construction &  Disabling Condition   \\
\hline
\pnt\ $\times$ \pnt\ $\rightarrow$ \lin\ & equal                          \\
\pnt\ $\times$ \lin\ $\rightarrow$ \pln\ & on                             \\
\pnt\ $\times$ \pln\ $\rightarrow$ \lin\ & (none)                         \\
\lin\ $\times$ \lin\ $\rightarrow$ \lin\ & parallel or intersect          \\
\lin\ $\times$ \pln\ $\rightarrow$ \pnt\ & in                             \\
\pln\ $\times$ \pln\ $\rightarrow$ \lin\ & parallel                       \\
 \hline
\end{tabular}
\caption{Operations for all unordered pairs formed from \pnt, \lin\ and \pln.
The operations cannot be performed if the condition listed holds between
the input arguments of the construction.
}
\label{pointsLinesPlanes}
\end{center}
\end{figure*}

Two of the constructions involve perpendiculars.
The line determined by the point
and the plane is the perpendicular
to the plane through the point.
The line determined by two lines in general position
likewise is a perpendicular:
the unique one that is perpendicular to both given lines.

In this way, an object-oriented reading
of Hilbert's axioms determines that the class \lin\
contains constructors with parameters
(\pnt, \pnt), with (\pnt, \pln), and with (\pln, \pln).
The class \pnt\ contains a constructor
with parameters (\lin, \pln).
The class \pln\ contains a constructor
with arguments (\pnt, \lin).

The constructors cannot be invoked
when the conditions noted in Figure~\ref{pointsLinesPlanes} hold
between the arguments.
For example, if a \pnt\ is on a \lin, then these do not determine a plane.
These conditions are semi-decidable: they either determine that the
condition does not hold, or that the condition \emph{may} hold.
However, in rare cases it can be determined that, say, two
instances of \pnt\ are equal.
The conditions therefore return the truth values of a 3-valued logic.

The abstract nature of an axiomatic approach to geometry
requires that the \pnt, \lin\ and \pln\ 
are left undefined.
This abstraction is not only essential in 
the axiomatic treatment of mathematical theories,
but it is also the essence of object-oriented design.

In object-oriented design one may distinguish two forms of abstraction.
The weaker form is achieved by any class in which the variables
are private.
One can then modify the representation of the objects without
consequences for the code using the class.
There is also a stronger form of abstraction in which
polymorphism makes it possible
to use more than one implementation of the same abstraction simultaneously.
The concept is then represented by an abstract class for the concept in
which the representation-dependent methods are virtual.
For each representation there is a separate derived class of
which the methods are dispatched at run time.
We have found this stronger form of abstraction
advantageous in our suite of \cpp classes.

\begin{figure}[!htbp]
\begin{center}
\epsfxsize=3.5in
\leavevmode
\epsfbox{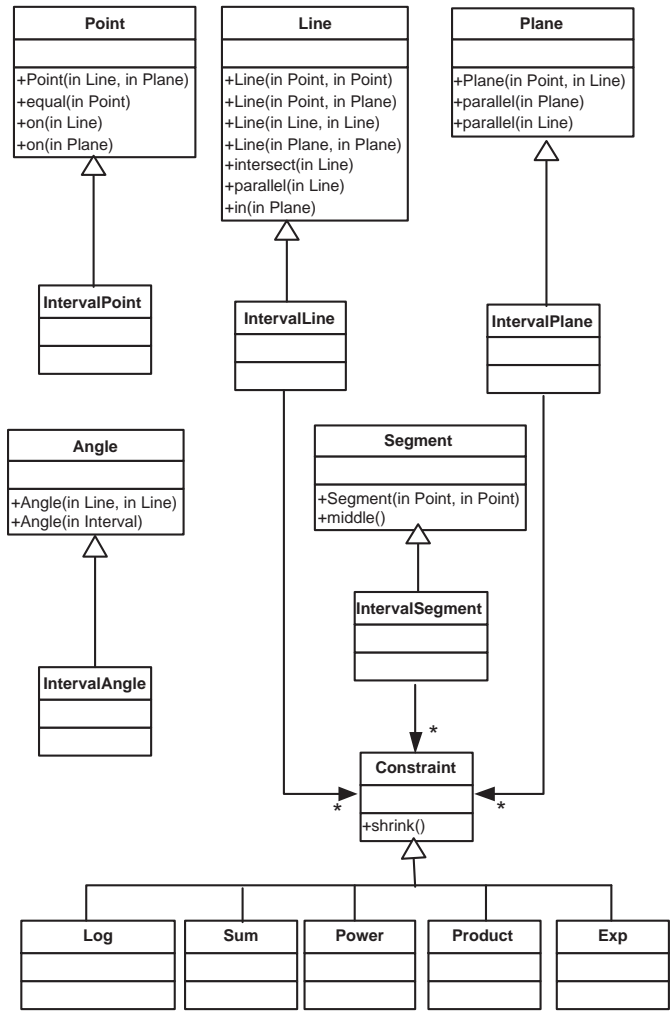}
\caption{
\label{geomuml}
UML class diagram for our system.}
\end{center}
\end{figure}

A UML class diagram summarizing the classes 
and the conditions above is shown in Figure~\ref{geomuml}. 
The purpose of the extra classes in the diagram is explained later.

\section{Consequences of computability limitations}

Whatever computer representation is chosen,
there will only be finitely many points, lines, and planes
that can be represented.
The conventional method of mapping
the infinity of abstract objects
to the finitely many representable ones
is to choose a representation in terms of reals
and then to map each real to a nearby floating-point number.
When this method is followed,
it has so far not been found possible
to give precise meaning to the outcomes
of tests such as whether a point is on a line.
The outcomes have to be interpreted
as ``probably not'' and ``possibly'',
depending on whether the computed distance
(subject to an unknown error)
is greater than a certain tolerance.

It may seem that this degree of uncertainty
is inherent in the limitation
to a finite number of representations.
This is not case.
Even when restricted to floating-point numbers,
it is possible to represent the point $p$
by a set $P$ of points containing $p$;
likewise, the line $l$ can be represented by a set $L$.
These sets are specified in terms of floating-point numbers,
so there are only finitely many of these.
Because of this finiteness it is decidable
whether the set of points contains any
that is on any in the set of lines.
It may seem computationally formidable
to make such a determination.
Actually, the techniques of interval constraints
make this perfectly feasible \cite{hcqvn99},
and this is what we use.

If it is determined that no point in $P$ is on any line in $L$,
then it is clear that $p$ is not on $l$.
If, on the other hand, some point in $P$ is on some line in $L$,
this says nothing about  whether $p$ is on $l$.
However, if $P$ and $L$ are, in a suitable sense, small,
then it follows that $p$ is close to $l$.
It is this asymmetry that is a consequence of the fact that the test for a point
on a line can at best be a 
semi-decision algorithm.
Similarly, the other tests
in Figure~\ref{pointsLinesPlanes} are semi-decision algorithms.

It is worth mentioning that to cope with the computability limitations 
in the area of computational geometry, the \emph{exact geometric computing} 
paradigm was proposed \cite{yap95}. This paradigm encompasses all techniques 
for which the outcomes are correct. As shown in \cite{itar98}, 
interval arithmetic can be used to do exact geometric computing.
This paper is also classified under this paradigm.

\section{Our implementation}
In the previous section we explained the need for interval methods
to ensure that in most cases where a test should have a negative
outcome, this is indeed proved numerically.
Interval methods can do this in several ways.
In \cite{itar98}, Br{\"o}nnimann {\it et al.} used 
interval arithmetic to dynamically bound 
arithmetic errors when computing tests (i.e. to compute dynamic filters).
In our case, we use interval constraints
not only to compute tests but also to implement
geometrical constructions. 
This means that the representations
of \pnt, \lin, and \pln\ are in the form of constraint
satisfaction problems.
For example, a plane is represented by the constraint
$ax+by+cz+d=0$, where
$a$, $b$, $c$ and $d$ are real-valued constants and
$x$, $y$, and $z$ are real-valued variables.
Due to computability limitations
discussed in the previous section,
the coefficients $a$, $b$, $c$ and $d$
are implemented as floating-point intervals.
For each point with coordinates
in these intervals, the constraint has a different
plane as solution. In this way our concrete representation is
a set of planes in the abstract sense.
The reader may refer to the following papers 
\cite{rthpra01}, \cite{bhlgfg99}, \cite{hckmdw98}, 
and \cite{hckvnmdn01} for more information
on constraints, propagation algorithms, interval constraints,
correctness and implementation of interval constraints.

As shown in Figure~\ref{geomuml},
the abstract classes \pnt, \lin, and \pln\ are extended 
and modelled using intervals and constraints.
The abstract class \cnt\ represents the constraint class,
which can be extended to implement primitive constraints
such as {\tt Sum} and {\tt Prod}. 
Each of these primitive constraints has 
a \emph{domain reduction operator} (DRO), 
represented by {\tt shrink}() method,
which removes inconsistent values from 
the domains of the variables in the constraint.
The DROs of primitive constraints are computed based on 
interval arithmetic. As an example, 
the {\tt Sum} constraint defined 
by $x+y=z$ has the
following DRO
\begin{eqnarray}
&&X^{new}=X^{old} \cap (Z^{old} - Y^{old})\nonumber \\
&&Y^{new}=Y^{old} \cap (Z^{old} - X^{old})\nonumber \\
&&Z^{new}=Z^{old} \cap (X^{old} + Y^{old})\nonumber 
\end{eqnarray}
where the intervals $X^{old}$, $Y^{old}$, 
and $Z^{old}$ are the domains of $x$, $y$, and $z$
respectively before applying the DRO,
and $X^{new}$, $Y^{new}$ and $Z^{new}$ are the
domains of $x$, $y$, and $z$ respectively after applying the DRO.
For a non-primitive constraint, such as \lin\ and \pln, 
we first decompose it into primitive constraints and then
use the propagation algorithm to implement the {\tt shrink}() method.
A simple version of this algorithm is shown in Figure~\ref{gpa}.
\begin{center}
\begin{figure}
\begin{center}
\fbox{
\parbox{1 cm}{
\small{
\begin{tabbing}
make $A$ the set of primitive constraints;\\
while \=( $A \neq \emptyset$) $\{$\\  
         \>choose a constraint $C$ from $A$ and apply its DRO;\\
         \> if one of the domains becomes empty, then stop;\\
         \>add \=to $A$ all constraints involving variables whose\\
         \>\> domains have changed, if any;\\
         \>remove $C$ from $A$; $\}$
\end{tabbing}
}
}
}
\end{center}
\caption{
\label{gpa}
Propagation algorithm.}
\end{figure}
\end{center}
In what follows, we present some examples
in two dimensions illustrating the use of our implementation.
We ran the examples on a Pentium II machine with a CPU rate of 400 MHz, 
and with 128 MB of memory. 
\paragraph{Are two points in the same side of a line?}
Let $L$ be a line represented by $[2.0,2.5]*x - [0.5,1.0]*y=[1.0,1.05]$.
Let $P$ and $Q$ be the two points represented respectively by
$([0.0,0.0],[0.0,0.0])$ and $([0.5,0.5],[0.5,0.5])$.
The question we are interested in is to
determine whether $P$ and $Q$ are on the same side of $L$.
Using the function \emph{sameSide(Point, Point)},
which checks whether two points are in the same side of a line,
our system outputs the following results:
\begin{verbatim}
Duration (musec): 179
True: the points are in the same side
\end{verbatim}
This means that our system was able to prove that
$P$ and $Q$ are in the same side of $L$.

Now suppose that $Q$ is represented by $([1 , 1],[0.5 , 0.5])$.
In this case, our system returned the following output:
\begin{verbatim}
Duration (musec): 133
False: the points are not in the same side
\end{verbatim}
If, somehow, the point $Q$ is only known to be represented by
$([0.75 , 1],[0.25 , 0.5])$ (note that the intervals are not singletons),
then our system was not able to prove that
the points $P$ and $Q$ are in the same side.
The output in this case is:
\begin{verbatim}
Duration (musec): 265
Undetermined
\end{verbatim}
\paragraph{Circumcenter of a triangle}
Given three points $P$, $Q$ and $R$ represented respectively by
$([0.0,0.0],[0.0,0.0])$, $([1.0,1.0],[0.5,0.5])$ and $([0.5,0.5],[1.0,1.0])$
we wish to find the center of the circle passing through $P$, $Q$ and $R$.
This example is taken from \cite{fnpr02}.
Since this center is given by the intersection of $L_1$ and $L_2$,
where $L_1$ is the line that passes through the middle of the segment $PQ$ and is
perpendicular to the line passing through $P$ and $Q$, and $L_2$
is the line that passes through the middle of the segment $QR$ and
is perpendicular to the line passing through $Q$ and $R$.
Using the \emph{intersect (Line)} function that
checks whether a line intersects with another
line, our system returned the following output:
\begin{verbatim}
Duration (msec):  6
True: the lines intersect at
x = [0.41666666666666663 , 
     0.41666666666666669]
y = [0.41666666666666663 , 
     0.41666666666666669]
\end{verbatim}
We then checked whether the point
$(x,y)$
is on the line $L_3$
that passes through the middle of the segment $PR$
and is perpendicular to the line passing through $P$ and $R$.
The output of our system indicates it is possible:
\begin{verbatim}
Duration (musec): 112
Undetermined
\end{verbatim}

\section*{Acknowledgements}
This research was supported by the University of Victoria and by
the Natural Science and Engineering Research Council of Canada.

\end{document}